\documentclass[a4paper,10pt]{article}
\usepackage{graphics}

\begin{document}
\title{Variational Monte Carlo Study of\, $_{\Lambda\Lambda}^ 4{H}$}

\author{\textit {Bhupali Sharma}$^a$\thanks{Present address :Arya Vidyapeeth 
College,Gauhati University,Assam,India} \\
$^a$\textit{Department of Physics, Jamia Millia Islamia,}\\
\textit{ New Delhi-110025, INDIA}\\}

\date{}
\maketitle
\begin{abstract}
Variational Monte Carlo calculations are carried out for 
$_{\Lambda\Lambda}^ 4{H}$ to explore on the possibility of its existence,
using realistic $NN$, $NNN$, and phenomenological $\Lambda N$ and 
$\Lambda NN$ interactions. We also perform calculations using FGNNand NSC97
$\Lambda N$ potentials. We have demonstrated the effect of the exchange part of the two-body $\Lambda N$ interaction and three-body $\Lambda NN$ interaction 
on the binding energy of $_{\Lambda\Lambda}^ 4{H}$. Still the stability of 
$_{\Lambda\Lambda}^ 4{H}$ remains an open question and this hypernuclear system has to be explored more, theoretically.\\
\end{abstract}

PACS numbers: 21.80.+a;21.60.Gx; 21.45.-v; 21.30.-x

\pagebreak

After the identification of  $_{\Lambda\Lambda}^6{He}$ event in E373 at KEK [1] ($\Delta B_{\Lambda\Lambda}= 1.01\; \pm\; 0.20^{+0.18}_{-0.11}$ MeV), the double-hypernuclear Physics  attained new horizons. This event then  became a testing ground for $\Lambda N $ and $\Lambda\Lambda$ potentials [2-6]. The first report of 
probable $_{\Lambda\Lambda}^4{H}$ hypernuclei from BNL, AGS experiment E906[7] 
has created theoretical interest on the possibility of existence of 
$_{\Lambda\Lambda}^4{H}$. 
A stable $_{\Lambda\Lambda}^4{H}$ system was predicted by Faddeev calculation 
in a $\Lambda\Lambda d$ model done by Filikhin and Gal [3] , with 
$B_{\Lambda\Lambda}\approx$ 0.3 MeV but their  four-body $\Lambda\Lambda p n$ 
exact Faddeev-Yakubovsky 
solution , for a wide range of $\Lambda\Lambda$ 
interaction strengths including the one that reproduces the $B_{\Lambda\Lambda}$ of $_{\Lambda\Lambda}^6{He}$, fails to bind $_{\Lambda\Lambda}^4{H}$ . 
Four-body stochastic variational model calculation with the correlated Gaussian
basis functions, by Nemura, Akaishi and Myint[4] has reported a bound state for
$_{\Lambda\Lambda}^4{H}$ with $B_{\Lambda\Lambda}$ $\approx$ 0.4 Mev. A variational Monte Carlo calculation for the binding energy $B_{\Lambda\Lambda}$ of the
lightest hypernuclei $_{\Lambda\Lambda}^4{H}$ was performed by M. Shoeb[5]
in the four body $\Lambda\Lambda p n$ model. A range of input 
$\Lambda\Lambda $ potentials of moderate strength could produce a particle 
stable $_{\Lambda\Lambda}^4{H}$ for the simulated NSC97e and NSC97f 
$\Lambda N$ potentials whereas the phenomenological Minnesota $\Lambda N$ 
potential seemed to need much stronger $\Lambda\Lambda $
 potential to bind. In our study we have performed variational calculations on
$_{\Lambda\Lambda}^4{H}$ to explore the possibility of its existence. The 
$\Lambda\Lambda$ interaction is the one reliably obtained by us earlier[6] 
performing complete six-body variational Monte-Carlo calculations for
$_{\Lambda\Lambda}^6{He}$ using realistic interactions with highly flexible 
correlations. Recently, the results of KEK E373 have been updated ( $\Delta B_{\Lambda\Lambda} = 0.67 \pm 0.17 $ MeV)[8]. Our $\Lambda\Lambda$ interaction is based on the earlier results of KEK E373[1].

From our calculations we find that the stability of 
$_{\Lambda\Lambda}^4{H}$ depends on the three-body $\Lambda $NN potential and 
the exchange part of the $\Lambda $N potential.

In the nuclear part of the Hamiltonian, we have used \textit{Argonne $V_{18}$} NN and \textit{Urbana IX} NNN potential same as ref.[6].

The phenomenological $\Lambda N$ potential consisting of central, Majorana spaceexchange and spin-spin $\Lambda N$ components is given by,

\begin{equation}
V_{\Lambda N} = (V_c(r) - \bar{V}T_{\pi}^2(r)) (1 - \epsilon+
 \epsilon P_x)
 + {{1}\over{4}} V_{\sigma}T_{\pi}^2(r)\sigma_{\Lambda}\cdot\sigma_{N}
\end{equation}

where $ P_x$ is the Majorana space-exchange operator and $\epsilon$ is the exchange parameter which we take as $0.2$[9].

This is consistent slightly on the low side with the forward to backward ratio of the low energy $\Lambda$p scattering data[10]. $V_c(r)$, $\bar{V}$ and $V_{\sigma}$ are respectively Wood-Saxon core, spin-average and spin-dependent strength and $ T_{\pi}^2(r)$ is one-pion tensor shape factor.

The $\Lambda NN $ potential consists of a two-pion exchange and a dispersive part[9] which arise mainly from elimination of $\Sigma$ degrees of freedom.

The two phenomenological forms for the dispersive kind are, the 
spin independent kind, $V_{\Lambda NN}^D$ and the spin-dependent kind,
$V_{\Lambda NN}^{DS}$. These are
 given by

\begin{equation}
V_{\Lambda NN}^D(r_{ij\Lambda}) = W_0 T_{\pi}^2(r_{i\Lambda}) 
T_{\pi}^2(r_{j\Lambda}) ,
\end{equation}

\begin{equation}
V_{\Lambda NN}^{DS} = V_{\Lambda NN}^D(r_{ij\Lambda}) [ 1 + {{1}\over{6}}
\vec\sigma_{\Lambda}\cdot(\vec\sigma_i + \vec\sigma_j)  ] . 
\end{equation}

 The two-pion exchange part of the interaction is given by[11] 

\begin{equation}
W_p = - {{1}\over{6}}C_p(\vec\tau_i\cdot\vec\tau_j)\{X_{i\Lambda},
X_{j\Lambda}\}Y_{\pi}(r_{i\Lambda})Y_{\pi}(r_{j\Lambda}) ,
\end{equation}

 where $ X_{k\Lambda}$ is the one-pion-exchange operator given by 

\begin{equation}
X_{k\Lambda} = (\vec\sigma_k\cdot\vec\sigma_{\Lambda}) + S_{k\Lambda}
(r_{k\Lambda})T_{\pi}(r_{k\Lambda})  
\end{equation}

 with 

\begin{equation}
S_{k\Lambda}(r_{k\Lambda}) = {{3(\vec\sigma_k\cdot
 r_{k\Lambda})(\vec\sigma_{\Lambda}\cdot r_{k\Lambda})}\over{r^2_{k\Lambda}}} -
 (\vec\sigma_k\cdot\vec\sigma_{\Lambda}) . 
\end{equation}

$Y_{\pi}(r_{k\Lambda})$ and $T_{\pi}(r_{k\Lambda})$ are the usual Yukawa and 
tensor functions with pion
 mass, $\mu$ = 0.7 ${fm}^{-1}$

Here, $C_p$ and $W_0$ are $\Lambda NN$ interaction parameters

The $\Lambda N$ and $\Lambda NN$ potential parameters for our three preferred models[6] are listed in Table I. $C_p$ and $W_0$ are the strength parameters of the two-pion and dispersive parts of the $\Lambda NN$ potential.

\vspace{1.5cm}

TABLE I : ${\Lambda}N$ and ${\Lambda}NN$ interaction parameters.
Except for $\epsilon$, all other quantities are in MeV.

\vspace{1cm}

\begin{quote}
\begin{tabular}{llllll}

\hline
\hline
&\\
$\Lambda N$ &$\bar V$ &$V_\sigma$&$\epsilon$&$C_p$&$W_0$\\
&\\

\hline
\hline
&\\
$\Lambda N1$&6.150&0.176&0.2&1.50&0.028\\
$\Lambda N2$&6.110&0.000&0.0&1.50&0.028\\
$\Lambda N3$&$6.025$&0.000&0.0&0.00&0.000\\
&\\
\hline
\hline
\vspace{7mm}
\end{tabular}
\end{quote}

For our study, we use low-energy phase equivalent Nijmegen interactions as our 
 $\Lambda\Lambda$ potential, which is represented by  the sum of three Gaussians [12,13,14] given by,

\begin{equation}
 V_{\Lambda \Lambda}  =  v^{(1)}exp(-{r^2}/\beta^2_{(1)}) +
\gamma v^{(2)} exp(-{r^2}/\beta^2_{(2)})  +  v^{(3)} exp(-{r^2}/
 \beta^2_{(3)}) 
\end{equation}

 where the strength parameter  $v^{(i)}$ and the range parameter
$\beta^{(i)}$ are taken   from [3]. For completeness, they
are displayed in Table II.
 The values of  $\gamma$ = 0.5463, 1.0 and
1.2044 correspond to   Nijmegen interactions NSC97e[13],
ND(NHC-D)[15], and NEC00(ESC00 or   NSC00)[16], respectively.

\vspace{1.5cm}
TABLE II : ${\Lambda\Lambda}$ interaction parameters.

\begin{quote}
\begin{tabular}{lll}
\hline
\hline
&\\
&$\beta_{(i)}(fm)$&$v^{(i)}(MeV)$\\
&\\
\hline
\hline
&\\
1.&1.342&-21.49\\
2.&0.777&-379.1\\
3.&0.350&9324\\
&\\

\hline
\hline

\end{tabular}
\end{quote}
\vspace{1cm}

This $ \Lambda \Lambda$ potential is a soft-core type which was fitted to Nijmegen
model   D by Hiyama \textit{et. al.}[14]. The short range term ($i=3$)
provides for a strong
soft-core repulsion and the long-range term ($i=1$) for attraction. The
 parameter
$\gamma$ which controls the strength of the mid-range attractive term ($i=2$)
 is
chosen such that the potential given by eq(2) reproduces the
 scattering
length and the effective range for a given model as close as possible.
Though, fully coupled-channel approach to doubly strange s-shell hypernuclei[17], predicted particle stable bound state of $_{\Lambda\Lambda}^4{H}$, 
in our potential the coupling between the different channels
($\Lambda\Lambda$,$\Xi N$,$\Sigma \Sigma$) have not been considered explicitly. 
The $\Lambda N$-$\Sigma N$ coupling and the $\Lambda \Sigma$ coupling for light hypernuclei has been taken care of by the three-body $\Lambda NN$ potantial. The
$\Lambda\Lambda$-$\Xi N$-$\Sigma \Sigma$ coupling for light hypernuclei is negligible[18] and hence is not considered here.

We take the variational wave function to be  of the form,

\begin{equation}
 \vert\Psi_v (\Lambda\Lambda)\rangle = [1 +\displaystyle
\sum_{i<j<k} (U_{ijk} + U^{TNI}_{ijk} )
 +\displaystyle\sum_{i<j,\Lambda} U_{ij,\Lambda} +
  \displaystyle\sum_{i<j} U_{ij}^{LS}] 
  \vert\Psi_p (\Lambda\Lambda)\rangle
\end{equation}

 where the pair wave function, $\vert\Psi_p (\Lambda\Lambda)\rangle$ is

\begin{equation}
 \vert\Psi_p (\Lambda\Lambda)\rangle = S \displaystyle\prod_
 {i<j} ( 1 + U_{ij} ) S\displaystyle\prod_{i<\Lambda} ( 1 + U_{i\Lambda} ) 
 \vert\Psi_J(_{\Lambda\Lambda}^A Z)\rangle
\end{equation}

The operator $S$ symmetrizes the various non-commuting operators which
occur in U.

The Jastrow wave function $\vert\Psi_J(_{\Lambda\Lambda}^A Z)\rangle$
for the s-shell $\Lambda\Lambda$ hypernucleus with two- and three- body central 
correlations represented by various fs is,

\begin{equation}
\vert\Psi_J(_{\Lambda\Lambda}^A Z)\rangle = f_c^{\Lambda\Lambda}
\displaystyle\prod_ {i<\Lambda}f_c^{i\Lambda}
\displaystyle\prod_ {i<j<k}f_c^{ijk}
\displaystyle\prod_ {i<j}f_c^{ij}
\vert\Psi_{JT}(^{A-2} Z)\rangle
A\vert\downarrow\Lambda\uparrow\Lambda\rangle
\end{equation}

where$\vert\Psi_{JT}(^{A-2} Z)\rangle$
represents the spin- and isospin wave-function of the s-shell nucleus with
definite total angular momentum J and isospin T 
 and 
$A\vert\downarrow\Lambda\uparrow\Lambda\rangle$ represents the anti-symmetric wave-function of the two $\Lambda$-particles coupled to total angular momentum zero. Correction to different $f$s are made by introducing cosine correction term[6] given as,

\begin{equation}
f\rightarrow f+\displaystyle\sum_{n=1}^4a_n cos\left({{n\pi r}\over {r_d}}
\right) for r\le{r_d}  
\end{equation}

where $a_n$ are variational parameters. The healing distance $r_d$ is also a variational parameter.

 First we perform calculations with the
potential models $\Lambda N1$ and $\Lambda N2$[6] and the
 results for calculation of binding energy of $_{\Lambda\Lambda}^4{H}$ 
are presented in Table III. The energies for $^2{H}$ and $^3{H}$ 
 are taken to be -2.2246(00) MeV and -8.32(01) MeV respectively. 
All the values in Table III are in MeV and SEC denotes space-exchange contribution.

\vspace{1cm}

TABLE III : Variational results for
$_{\Lambda\Lambda}^4{H}$ and $_{\Lambda}^3{H}$ 
with $\Lambda N1$ and $\Lambda N2$ 

\vspace{1 cm}

\begin{quote}
\begin{tabular}{llll}

\hline
\hline

&\\
Potential&Terms &$_{\Lambda\Lambda}^4{H}$&$_{\Lambda}^3{H}$\\
&\\
\hline
\hline
&\\
$\Lambda N1$& E &-2.60(03) & -2.56(01)\\
&SEC&0.37(00)&0.04(00)\\
&${B_{\Lambda\Lambda}}$& {0.38(03)}&\\
&${B_{\Lambda}}$&& {0.34(01)} \\
&\\
$\Lambda N2$&E & -2.53(00) & -2.35(00)\\
&SEC&0.00(00)&0.00(00)\\
&${B_{\Lambda\Lambda}}$&{0.31(00)}&\\
&${B_{\Lambda}}$&&{0.13(00)}\\
&$B_{\Lambda}$(expt.)&& 0.13$\pm$ 0.05 [19]\\
&\\
\hline
\hline

\end{tabular}
\end{quote}

\vspace{1.5 cm}

These results show that for the potential model $\Lambda N1$, for which the
space exchange parameter $\epsilon$=0.2,
$_{\Lambda\Lambda}^4{H}$ is with respect to the deuteron energy, but is unstable with respect to the $_{\Lambda}^3{H}$ energy.
For the potential
model $\Lambda N2$, for which $\epsilon$=0, it is well bound and is consistent with $_{\Lambda}^3{H}$ energy.
For the third potential model $\Lambda N3$, for which $\Lambda NN$
interaction is absent, $_{\Lambda\Lambda}^4{H}$
was found to be unbound. This indicates that
the effect of space exchange parameter and $\Lambda NN$ force is
important for $_{\Lambda\Lambda}^4{H}$. Previous studies on
$_{\Lambda\Lambda}^4{H}$ are subject to debate due to contradictory results.
Faddeev calculations by Filikhin and Gal[12,20] predicted a bound
$_{\Lambda\Lambda}^4{H}$ with $B_{\Lambda\Lambda}$$\approx$0.3 MeV using
$\Lambda\Lambda d$ model whereas their $\Lambda\Lambda pn$ Faddeev-
Yakubovsky calculations found $_{\Lambda\Lambda}^4{H}$ to be unbound. Again,
Nemura \textit{et. al.}[4] and M. Shoeb[5] found $_{\Lambda\Lambda}^4{H}$
to be bound with $B_{\Lambda\Lambda}$$\approx$0.4 MeV.

Therefore we performed further
calculations on
$_{\Lambda\Lambda}^4{H}$ using various kinds of interactions. 

First we performed calculations on $_{\Lambda\Lambda}^4{H}$, $_{\Lambda}^3{H}$,
$_{\Lambda}^4{H}$ and $_{\Lambda}^4{H}^*$ by  taking
$\gamma=0.681$[6], $\epsilon=0$, $C_p=0$,
 $W_0=0$ (same  as in $\Lambda N3$)  and having different
  sets of values of $\bar{V}$ ans $V_{\sigma}$. These results are
  tabulated in Table IV. We chose the values of $\bar{V}$
  and $V_\sigma$ in such a manner that $_{\Lambda}^3{H}$ is bound in
  each case. These  calculations were done to search a bound
  $_{\Lambda\Lambda}^4{H}$ with two-body central  $\Lambda N$
  potential. All the energy values are in MeV and the distances are in fm.

\vspace{1cm}

TABLE IV : Variational results for
$_{\Lambda\Lambda}^4{H}$, $_{\Lambda}^3{H}$,  $_{\Lambda}^4{H}$
$_{\Lambda}^4{H}^*$  with  different values of
$V_\sigma$ and $\bar{V}$

\vspace{1.5 cm}

\begin{tabular}{lp{0.5cm}llll}

\hline
\hline

&\\
&Term &$_{\Lambda\Lambda}^4{H}$&$_{\Lambda}^3{H}$&$_{\Lambda}^4{H}$&
$_{\Lambda}^4{H}^*$\\
&\\
\hline
\hline

$V_\sigma=0,$  & E & -2.83(00)&-2.35(00)&-11.06(01)&-11.06(01)\\
$\bar{V}=6.25$&${B_{\Lambda\Lambda}}$&{0.61(00)}&&&\\
&${B_{\Lambda}}$&&{0.13(00)}&
{2.74(01)}&{2.74(01)}\\
&\\
$V_\sigma=0.1,$  & E & -2.64(01)&-2.37(00)&-10.87(01)&-10.59(01)\\
$\bar{V}=6.21$&${B_{\Lambda\Lambda}}$&{0.42(01)}&&&\\
&${B_{\Lambda}}$&&{0.15(00)}&
{2.55(01)}&{2.27(01)}\\
&\\

$V_\sigma=0.15,$  &E & -2.44(00)&-2.37(00)&-10.78(01)&-10.33(01)\\
$\bar{V}=6.19$&${B_{\Lambda\Lambda}}$&{0.22(00)}&&&\\
&${B_{\Lambda}}$&&{0.15(00)}&{
2.46(01)}&{2.01(01)}\\
&\\

$V_\sigma=0.2,$  &E & -2.40(01)&-2.38(00)&-10.69(01)&-10.10(01)\\
$\bar{V}=6.17$&${B_{\Lambda\Lambda}}$&{0.18(01)}&&&\\
&${B_{\Lambda}}$&&{0.16(00)}&
{2.37(01)}&{1.78(01)}\\
&\\
$V_\sigma=0.25,$  &E & -2.30(00)&-2.37(00)&-10.54(01)&-9.86(01)\\
$\bar{V}=6.14$&${B_{\Lambda\Lambda}}$&{0.08(00)}&&&\\
&${B_{\Lambda}}$&&{0.15(00)}&
{2.22(01)}&{1.54(01)}  \\
&\\

$V_\sigma=0.3,$ &E & -2.22(00)&-2.36(01)&-10.44(01)&-9.58(01)\\
$\bar{V}=6.12$&${B_{\Lambda\Lambda}}$&{-0.00(00)}&&&\\
&${B_{\Lambda}}$&&{0.14(01)}&
{2.12(01)}&{1.26(01)} \\
&\\

$V_\sigma=0.4,$ &E & -2.16(00)&-2.35(01)&-10.19(01)&-9.12(01)\\
$\bar{V}=6.07$&${B_{\Lambda\Lambda}}$&{-0.06(00)}&&&\\
&${B_{\Lambda}}$&&{0.13(01)}&
{1.87(01)}&{0.80(01)} \\
&${B_{\Lambda}}$(expt.)&&0.13$\pm$0.05[19]&-10.36(04)[*]&-9.32(06)[*]\\

\hline
\hline
\end{tabular}

\vspace{1cm}

(* calculated by taking experimental ${B_{\Lambda}}$ and ${B_{\Lambda}^*}$ values 2.04(04)[19] and 1.04(04)[21] and the energy for $^3 H$ to be -8.32(01).)

The $V_\sigma$ vs -E graph for $_{\Lambda\Lambda}^4{H}$,
$_{\Lambda}^4{H}$ and $_{\Lambda}^4{H}^*$ is shown in figure 1.
In the figure the energies for $_{\Lambda\Lambda}^4{H}$ are taken as
-(E + 6), for the sake of comparison  with the corresponding values of
$_{\Lambda}^4{H}$ and $_{\Lambda}^4{H}^*$. For the experimental values of
energy for
$_{\Lambda}^4{H}$ and $_{\Lambda}^4{H}^*$, we take the values of $B_\Lambda$
and $B_\Lambda^*$ from Ref.[18,19] and energy value for $^3H$ to be -8.32(01)[9].
 It is seen from the figure that
corresponding to the experimental values of energy for
$_{\Lambda}^4{H}$ and $_{\Lambda}^4{H}^*$
,-10.36(04)  and -9.32(06) respectively (calculated by taking experimental $B_{\Lambda}$ and $B_{\Lambda}^*$ values 2.04(04) and 1.00(06))[18,19],  $_{\Lambda\Lambda}^4{H}$,
is  unbound.

\vspace{2cm}

\begin{figure}[h]
\vspace{1cm}
\begin{centering}
\resizebox{8cm}{!}{\includegraphics{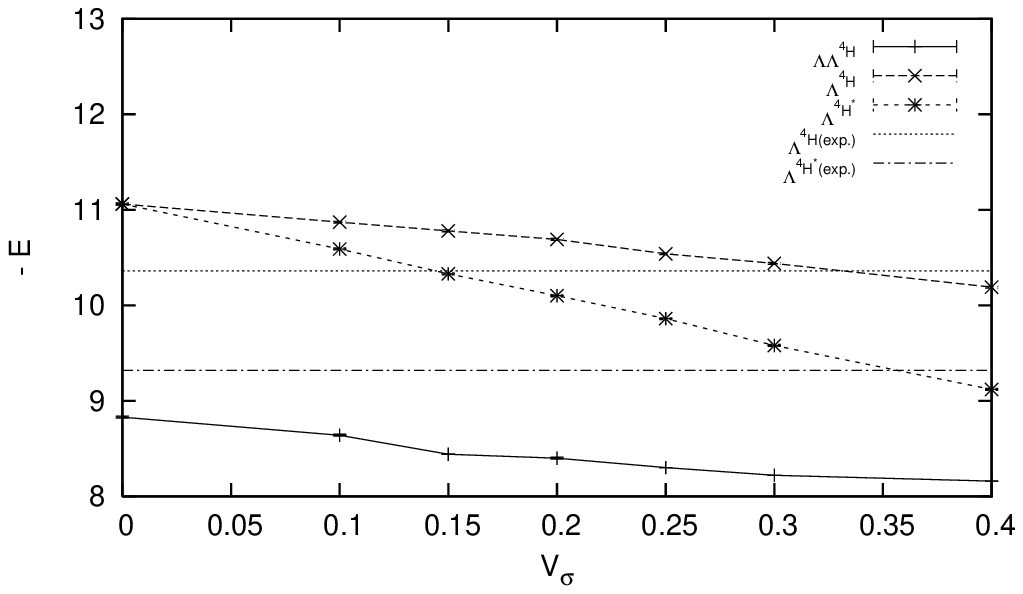}}

\caption{ $V_\sigma$ vs -E graph}
\end{centering}
\end{figure}

\vspace{1cm}

Then we performed calculations for different Nijmegen interactions with
different values of $\gamma$ with
 $\epsilon$=0, $C_p$=0 and $W_0$=0. These results are given in
  Table V. For $\gamma$=0.773, which is intermediate between NSC97e[11] and ND[13], we call the interaction as NM[6]. 
The   energy value for $^2 H$ for calculation of $B_{\Lambda\Lambda}$ is taken
to be -2.2246(00).

\pagebreak

TABLE V : Variational results for
$_{\Lambda\Lambda}^4{H}$ with different Nijmegen interactions

\begin{quote}
\begin{tabular}{lllrrr}

\hline
\hline

&\\
Potential&$\gamma$&Term &$V_\sigma=0.2$&$V_\sigma=0.3$&$V_\sigma=0.4$\\
&&&$\bar{V}$=6.17&$\bar{V}$=6.12&$\bar{V}$=6.07\\
&\\
\hline
\hline
&\\
NSC97e& 0.5463& E & -2.32(00)&-2.21(00)&-2.09(00)\\
&&${B_{\Lambda\Lambda}}$&{0.10(00)} & {-0.01(00)}&{-0.13(00)} \\
&\\
NM& 0.773& E & -2.53(01)&-2.28(00)&-2.19(01)\\
&&${B_{\Lambda\Lambda}}$&{0.31(01)} & {0.06(00)}&{-0.03(01)}\\
&\\
ND& 1.0& E & -2.89(01)&-2.30(00)&-2.23(00)\\
&&${B_{\Lambda\Lambda}}$&{0.67(01)} &{ 0.08(00)}& {0.01(00)}\\
&\\
NEC00& 1.2044& E & -3.34(01)&-2.35(00)&-2.29(01)\\
&&${B_{\Lambda\Lambda}}$&{1.12(01)}&{0.13(00)} &{0.07(01)} \\

&\\

\hline
\hline

\end{tabular}
\end{quote}

\vspace{1cm}

The $\gamma$ vs -E graph for $_{\Lambda\Lambda}^4H$ is shown in figure 2.
The graph shows that -E increases with $\gamma$
which is due to the fact that as $\gamma$ increases,
$\Lambda\Lambda$ potential becomes more attractive. From this
graph also it is seen that for $V_\sigma$=0.3-0.4, which includes the
experimental values for $_{\Lambda}^4H$ and $_{\Lambda}^4H^*$[16,17],
$_{\Lambda\Lambda}^4H$ is not bound.

\vspace{1cm}

\begin{figure}[h]
\begin{centering}
\resizebox{8cm}{!}{\includegraphics{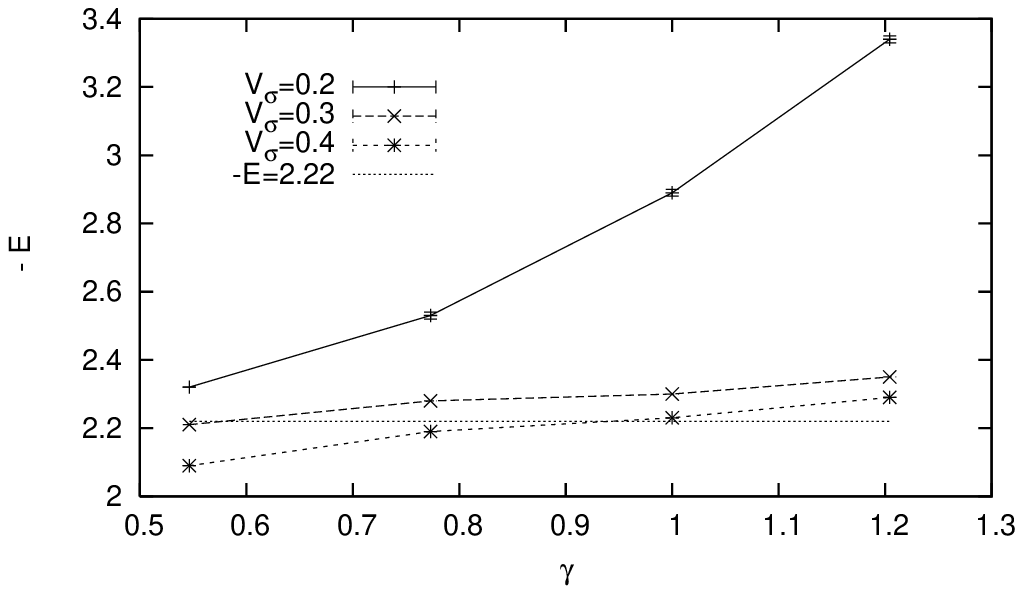}}

\caption{ $\gamma$ vs -E graph for $_{\Lambda\Lambda}^4H$}
\end{centering}
\end{figure}
\vspace{1cm}

Since Nemura \textit{et. al}[4] and
M. Shoeb[5] had predicted a bound $_{\Lambda\Lambda}^4{H}$,
 we performed calculations with NN potential used by Filikhin and Gal(FGNN)[3]
 and NSC97 $\Lambda$N potential[3], the same as Nemura \textit{et. al}[4] and
M. Shoeb[5] with which they found bound $_{\Lambda\Lambda}^4{H}$.
These results are given in Table VI. The corresponding results for 
Argonne $v_ {18}$ NN potential(same as ref.[6])  and NSC97 $\Lambda$N[3]
 potential are given in Table VII for the hypernuclear systems, 
$_{\Lambda}^4{H}$ and $_{\Lambda}^4{H}^*$. In the following tables, all energy 
values are in MeV and all distances are in fm.

\vspace{2cm}

TABLE VI : Variational results for $_{\Lambda}^3{H}$ and
$_{\Lambda\Lambda}^4{H}$ with $\epsilon$=0,  FGNN
potential and NSC97 $\Lambda$N potential

\vspace{1cm}

\begin{quote}
\begin{tabular}{p{3cm}llll}

\hline
\hline

&\\
&&Term &$_{\Lambda}^3{H}$&$_{\Lambda\Lambda}^4{H}$\\
&\\
\hline
\hline
&\\
NSC97e&& E & -2.37(02)&-2.57(02)\\
&&${B_{\Lambda}}$&{0.15(02)}&\\
&&${B_{\Lambda\Lambda}}$&&{0.35(02)} \\
&\\
&Ref.[5]&${B_{\Lambda}}$&{0.018}  &\\
&&${B_{\Lambda\Lambda}}$&&{0.37}\\
&\\
&\\
NSC97f&& E & -2.47(02)&-2.58(03)\\
&&${B_{\Lambda}}$&{0.25(02)}&\\
&&${B_{\Lambda\Lambda}}$&&{0.36(03)}\\
&\\
&Ref[4]&${B_{\Lambda}}$&{0.24}\\
&&${B_{\Lambda\Lambda}}$&&{0.4}\\
&Ref.[5]&${B_{\Lambda}}$&{0.26}\\
&&${B_{\Lambda\Lambda}}$&&{0.41}\\
&&$B_{\Lambda}$(expt.)& 0.13$\pm$ 0.05 [19]&\\

&\\

\hline
\hline

\end{tabular}
\end{quote}

\pagebreak

TABLE VII  : Variational results for $_{\Lambda}^3{H}$,
$_{\Lambda\Lambda}^4{H}$, $_{\Lambda}^4{H}$ and $_{\Lambda}^4{H}^*$
with  Argonne $V_{18}$ NN and NSC97
$\Lambda$N potential

\vspace{1cm}

\begin{quote}
\begin{tabular}{llllll}
\hline
\hline

&\\
Potential&Term &$_{\Lambda}^3{H}$&$_{\Lambda\Lambda}^4{H}$&
$_{\Lambda}^4{H}$&$_{\Lambda}^4{H}^*$\\
&\\

\hline
\hline

&\\
NSC97e& E & -2.32(01)&-2.51(02) & -10.57(03)&-10.38(02)\\
$\epsilon$=0&${B_{\Lambda}}$&{0.10(01)}&& 2.25(03)&2.06(02)\\
&${B_{\Lambda\Lambda}}$&&{0.29(02)} \\
&\\
NSC97e& E & -2.27(01)&-2.30(02)& -10.37(03)&-10.15(03)\\
$\epsilon$=0.2&${B_{\Lambda}}$&{0.05(01)}&&2.05(03)&1.83(03)\\
&${B_{\Lambda\Lambda}}$&&{0.08(02)}& \\
&\\
NSC97f& E & -2.43(01)&-2.54(02)&-10.84(03)&-10.38(03)\\
$\epsilon$=0&${B_{\Lambda}}$&{0.21(01)}&&2.52(03)&2.06(02)\\
&${B_{\Lambda\Lambda}}$&&{0.32(02)}& \\
&\\
NSC97f& E & -2.37(01)&-2.35(02)& -10.63(03)&-10.12(03)\\
$\epsilon$=0.2&${B_{\Lambda}}$&{0.15(01)}&&2.31(03)&1.80(03)\\
&${B_{\Lambda\Lambda}}$&&{0.13(02)}& \\
&${B_{\Lambda}}$(expt.)&0.13$\pm$0.05[19]&&-10.36(04)[*]&-9.32(06)[*]\\

&\\
\hline
\hline

\end{tabular}
\end{quote}

(* calculated by taking experimental ${B_{\Lambda}}$ and ${B_{\Lambda}^*}$ values 2.04(04)[19] and 1.04(04)[21] and the energy for $^3 H$ to be -8.32(01).)

\vspace{1cm}

From the results it has been observed that for FGNN potential
and NSC97 $\Lambda$N potential, with $\epsilon$=0 (same as Nemura
\textit{et. al.}[4] and M. Shoeb[5]), $_{\Lambda\Lambda}^4{H}$ is
 bound (Table VI). This result is similar to the results
of ref. [4] and [5]. For Argonne $V_{18}$ potential and
NSC97 $\Lambda$N potential (Table VII), with $\epsilon$=0,  $_{\Lambda\Lambda}^4{H}$
seems to be bound but for $\epsilon$=0.2, it is unbound.  Thus, the question
of the existence of $_{\Lambda\Lambda}^4{H}$ is seen to be dependent on the SEC in the  $\Lambda$N potential.

Our calculations indicate that the exchange part of two-body $\Lambda N$ interaction is very crucial for stability of $_{\Lambda\Lambda}^4{H}$.

Also, the three-body $\Lambda NN$ interaction seems to play a sensitive role towards the stability of $_{\Lambda\Lambda}^4{H}$.Since experimentally there is no concrete evidence of existence of $_{\Lambda\Lambda}^4{H}$, theoretically, the question of stability of $_{\Lambda\Lambda}^4{H}$ is to be explored more in future.

Bhupali Sharma is thankful to Prof. Q.N. Usmani and Prof. A.R. Bodmer for valuable guidance, to Dr. Tabish Qureshi for his help during the entire period of computation and also to the authorities of JMI for the computational 
facilities at the department of Physics, JMI.

\onecolumn

REFERENCES :

[1] H. Takahashi et. al. 2001 Phys. Rev. Lett 87 212502 
 
[2] I. N. Filikhin and A. Gal 2002 Nucl. Phys. A 707 491  

[3] I. N. Filikhin and A. Gal 2002 Phys. Rev. Lett. 89 172502  

[4] H. Nemura, Y. Akaishi and K. S. Myint 2003 Phys. Rev. C 67 051001(R) 

[5] M. Shoeb. 2004 Phys. Rev. C 69 054003

[6] Q. N. Usmani, A. R. Bodmer and Bhupali Sharma 2004 Phys. Rev. C 

70 061001(R)

[7] J. K. Ahn  et. al. 2001 Phys. Rev. Lett. 87 132504 

[8] K. Nakazawa 2010 Nucl. Phys. 835 207 ; K. Nakazawa and 

H. Takahashi 2010 Prog Theor Phys. Supplement 185 335 

[9] Rita Sinha, Q. N. Usmani and B. M. Taib 2002 Phys Rev C 66 024006 

[10] G. Alexander et. al. 1968 Phys. Rev. 173 1452 ; B. Sechi-Zorn et. al. 

1968 Phys. Rev. 175 1735 

[11] R.K. Bhaduri, B.A. Loiseau, and Y. Nogami 1967 Annals of Phys. 44 

57 

[12]I. N. Filikhin and A. Gal 2002 Phys. Rev. C 65 041001 

[13]V.G.J. Stokes and Th. A. Rijken 1999 Phys. Rev. C 59 3009 

[14]E. Hiyama et. al. 1997 Prog. Theor. Phys. 97 887 

[15]M. M. Nagels, T. A. Rijken, J.J. de Swart 1977 Phys. Rev. D 15 2547

[16] Th. A. Rijken 2001 Nucl. Phys.  A 691 322c 

[17] Nemura et. al. 2005 Phys. Rev. Lett. 94 202502 

[18] T. Yamada 2004 Phys. Rev  C 69 044301

[19] M. Juric et. al. 1973 Nucl. Phys. B52 1 

[20] I. N. Filikhin,  A. Gal and V. M. Suslov 2002 Phys. Rev. C 68

024002 

[21] M. Bedjidian et. al. 1979 Phys. Lett. 83 B 252

\end{document}